\documentclass[twocolumn,aps,prl,groupedaddress,showpacs]
              {revtex4}%
\usepackage{graphicx}
\usepackage{amsmath}
\usepackage{bm}
\usepackage{relsize}
\RequirePackage{xspace}
\newcommand{\babar}{\mbox{\slshape B\kern-0.1em{\smaller A}\kern-0.1em %
B\kern-0.1em{\smaller A\kern-0.2em R}}\xspace}

\begin{document}

\title{\mbox{}\\[10pt]
Inclusive Production of Four Charm Hadrons \\
in $\bm{e^+ e^-}$ Annihilation at $\bm{B}$ Factories}

\author{Daekyoung Kang, Jong-Wan Lee, and Jungil Lee}
\affiliation{
Department of Physics, Korea University, Seoul 136-701, Korea}

\author{Taewon Kim and Pyungwon Ko}
\affiliation{
Department of Physics, KAIST, Daejon 305-701, Korea}



\date{\today}
\begin{abstract}
Measurements by the Belle Collaboration of the
exclusive production of two charmonia in $e^+ e^-$ annihilation
differ substantially from theoretical predictions. 
Till now,
no conclusive explanation for this remarkable discrepancy has been
provided. Even the origin of the discrepancy is not identified, yet.
We suggest that the measurement of four-charm events in Belle
data must provide a strong constraint in identifying the origin of 
this large discrepancy. 
Our prediction of the cross section for $e^+e^-\to c\bar{c}c\bar{c}$, 
in lowest order in strong coupling constant,
at $\sqrt{s}=10.6$~GeV is about $0.1$~pb.
If measured four-charm cross section is compatible
with the prediction based on perturbative QCD, it is very likely that 
factorization of hadronization process from perturbative part may be
significantly violated or there exists a new production mechanism. 
If the cross section for the four-charm event 
is also larger than the prediction like that for the exclusive
$J/\psi+\eta_c$ production, perturbative QCD expansion itself will be proved
to be unreliable and loses predictive power.

\end{abstract}


\maketitle

The Belle Collaboration has measured the cross section for
$J/\psi + \eta_c$ by observing a peak in the momentum spectrum of the
$J/\psi$ that corresponds to the 2-body final state $J/\psi + \eta_c$
\cite{Abe:2002rb}. The measured cross section is
\begin{eqnarray}
\sigma[J/\psi+\eta_c] \times B^{\eta_c}[\ge 4]
= \left( 33^{+7}_{-6} \pm 9 \right) \; {\rm fb},
\label{Belle}
\end{eqnarray}
where $B^{\eta_c}[\ge 4]$ is the branching fraction for the $\eta_c$ to decay
into at least 4 charged particles.  Since $B^{\eta_c}[\ge 4]<1$, 
the right side
of Eq.~(\ref{Belle}) is a lower bound on the cross section to produce
$J/\psi + \eta_c$. This lower bound is about an order of magnitude
larger than the predictions by Braaten and Lee~\cite{Braaten:2002fi},
and by Liu, He, and Chao~\cite{Liu:2002wq} 
of nonrelativistic QCD(NRQCD)~\cite{BBL} in the nonrelativistic limit. 
The cross section was calculated previously by Brodsky and 
Ji~\cite{Brodsky:1985cr} using perturbative-QCD factorization 
formalism~\cite{Brodsky:1981kj}. But they did not give an 
analytic expression for the cross section. Recently, Brodsky, Ji, and
Lee redid~\cite{pQCD} the calculation 
given in Ref.~\cite{Brodsky:1985cr}. They
found exact agreement~\cite{pQCD} with the result
based on NRQCD~\cite{Braaten:2002fi,Liu:2002wq}. 
Currently, the cross section for the process $e^+e^-\to J/\psi+\eta_c$ 
shows the largest discrepancy between theory and data available 
within standard model.
This presents a challenge to our current understanding 
of charmonium production based on perturbative QCD framework.

\begin{figure*}
\includegraphics[width=11.5cm]{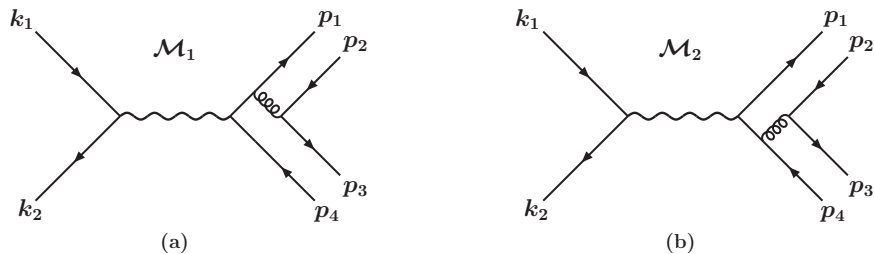}
\caption{\label{fig1}
Two of the 8 topologically distinct
Feynman diagrams for
$e^-(k_1) e^+ (k_2)\to c(p_1) \bar{c}(p_2) c(p_3) \bar{c}(p_4)$.
}
\end{figure*}

A few theoretical studies have been carried out in order to explain
the large discrepancy. Bodwin, Braaten, and Lee 
proposed~\cite{Bodwin:2002fk,Bodwin:2002kk} that the Belle
data for $J/\psi + \eta_c$ may include the $J/\psi+J/\psi$ events
because the width of the $\eta_c$ peak 
in the recoil mass distribution for inclusive $J/\psi$ production
measured by the Belle Collaboration
is wide enough to accommodate $J/\psi$ events. 
The large enhancement from photon fragmentation in the two-photon mediated
process $e^+e^-\to J/\psi+J/\psi$
overcomes the suppression factor $\alpha^2/\alpha_s^2$
in couplings compared to the 
$e^+e^-\to J/\psi + \eta_c$~\cite{Bodwin:2002fk,Bodwin:2002kk}.
Brodsky, Goldhaber, and Lee introduced an exotic scenario that
the Belle $J/\psi+\eta_c$ signal may include the associate 
production of $J/\psi$ 
and spin-$J$ glueball $\mathcal{G}_J$, $J=0,$ 2~\cite{Brodsky:2003hv}. 
The Belle Collaboration carried out an updated analysis~\cite{Abe:2004ww}
motivated by these proposals. According to the Belle analysis, no
events for $J/\psi+J/\psi$ have been detected and the upper limit for
$\sigma(e^+e^-\to J/\psi+ J/\psi)\times B^{J/\psi}[>2]$ is
$9.1$~fb at the 90\% C.L., which is consistent with the prediction given
in Refs.~\cite{Bodwin:2002fk,Bodwin:2002kk}. 
Here $B^{J/\psi}[> 2]$ is the branching fraction for the $J/\psi$ to decay
into more than 2 charged particles. 
The measured cross section for $e^+e^-\to J/\psi +J/\psi$,
however, does not explain the large fraction of the  $J/\psi + \eta_c$
signals. The Belle Collaboration also analyzed the angular distribution
of $J/\psi$ in order to identify if the data include the
associated $J/\psi+\mathcal{G}_0$ signals. The predicted angular 
distributions are proportional to $\cos^2\theta$ for 
$J/\psi+\eta_c$~\cite{Braaten:2002fi} and 
$J/\psi+\mathcal{G}_2$~\cite{Brodsky:2003hv}, 
and $\sin^2\theta$ for $J/\psi+\mathcal{G}_0$~\cite{Brodsky:2003hv}, 
where $\theta$ is the scattering angle of the $J/\psi$ in the 
$e^+e^-$ c.m. frame. The updated analysis show the measured distribution
is proportional to $\cos^2\theta$, which ruled out the spin-0 glueball
scenario~\cite{Abe:2004ww}. 
Since the angular distribution for $J/\psi+\mathcal{G}_2$
has the same form as that for $J/\psi+\eta_c$, spin-2 glueball has not
been ruled out, yet.

Another scenario is that higher-order corrections in strong coupling
$\alpha_s$ may be huge~\cite{Hagiwara:2003cw}. If it is true,
perturbative expansion is not a proper method to predict the cross section.
If it is not, it is probable that the factorization of long-distance
factor involving hadronization is seriously violated or there exists
a unknown production mechanism which we do not understand, yet.
Once we can estimate the size of the perturbative QCD corrections to
this process, it might be easier for us to identify the origin of
this large discrepancy. Unfortunately, the next-to-leading-order
corrections to the cross section of exclusive $J/\psi+\eta_c$ process
is not available. A comprehensive review on recent developments in 
quarkonium physics can be found in Ref.~\cite{Brambilla:2004wf}.

In this paper, we introduce an economical method
to check if the perturbative QCD corrections to the cross section  
is indeed large enough to explain the discrepancy. 
If we consider inclusive four-charm-hadron production, the prediction
for the cross section can be expressed as inclusive $c\bar{c}c\bar{c}$
production rate $\sigma(e^+e^-\to c\bar{c}c\bar{c}+X)$. 
This is analogous to estimating $\sigma(e^+e^-\to \textrm{hadrons})$
by $\sum_{q}\sigma(e^+e^-\to q\bar{q}+X)$. 
We expect $\sigma(e^+e^-\to c\bar{c}c\bar{c}+X)\approx%
\sigma(e^+e^-\to c\bar{c}c\bar{c})$ is  a good approximation
at $\sqrt{s}=10.6~$GeV. Unlike the prediction for $J/\psi+\eta_c$
cross section, the prediction for the inclusive four-charm-hadron 
production rate purely consists of short-distance factor. Corresponding
long-distance factor for hadronization is of order 1. Since this process
involves the same Feynman diagrams for exclusive $J/\psi+\eta_c$ 
production, the measurement of the cross section for four-charm-hadron 
production will provide an important information in estimating the size
of the short-distance coefficient for $J/\psi+\eta_c$ cross section. 
We present our prediction for inclusive four-charm-hadron production
by calculating  $\sigma(e^+e^-\to c\bar{c}c\bar{c}+X)$ in order 
$\alpha^2\alpha_s^2$, which is leading order in strong coupling constant.
If our leading-order prediction is comparable to the measured value,
it is very probable that the QCD higher-order corrections to the
$J/\psi+\eta_c$ cross section is small. Then the large discrepancy in
$J/\psi+\eta_c$ cross section may be due to the violation of factorization
or existence of new production mechanism. If the measured cross section
for the four-charm-hadron inclusive production is much larger than
our prediction like the case of $J/\psi+\eta_c$, it is very likely that
perturbative QCD corrections to $J/\psi+\eta_c$ cross section is 
large enough to explain the discrepancy, which leads to the failure 
of reliability in perturbative expansion. 

In leading order in strong coupling $\alpha_s$, 
$c\bar{c}c\bar{c}$ can be produced at order $\alpha^2\alpha_s^2$.
There are two topologically distinct Feynman diagrams generating two pairs of
$c\bar{c}$, which are shown as $\mathcal{M}_1$ and $\mathcal{M}_2$
in Fig.~\ref{fig1}(a) and \ref{fig1}(b), respectively.
Momenta for the involving particles are assigned as
$e^-(k_1) e^+ (k_2)\to c(p_1) \bar{c}(p_2) c(p_3) \bar{c}(p_4)$.
The amplitude for the two diagrams shown in Fig.~\ref{fig1} are
\begin{eqnarray}
-i\mathcal{M}_i&=&
i\frac{(4\pi)^2 e_c\alpha \alpha_s}{s(p_2+p_3)^2}
\bar{v}_e(k_2)\gamma_\alpha u_e(k_1)
\nonumber\\&\times&
\bar{u}(p_3)T^a\gamma_\beta v(p_2)
\,\bar{u}(p_1)T^a H_i^{\alpha\beta}v(p_4),
\label{eq:amp}
\end{eqnarray}
where $s=(k_1+k_2)^2$,
$e_c=\frac{2}{3}$ is the fractional electric charge of the
charm quark, and $a$ is the SU(3) color index for the virtual gluon.
The vector indices $\alpha$ and $\beta$ are for the virtual photon
and gluon, respectively. 
We suppress the spin and color indices of 
the charm quarks in Eq.~(\ref{eq:amp}).
For $i=1$ or 2 the tensors ${H}_i^{\alpha\beta}$ in Eq.~(\ref{eq:amp}),
which are matrices in spinor space, are defined by
\begin{subequations}
\begin{eqnarray}
H_1^{\alpha\beta}&=&\gamma^\beta\Lambda(p_1+p_2+p_3)\gamma^\alpha,
\\
H_2^{\alpha\beta}&=&\gamma^\alpha\Lambda(-p_2-p_3-p_4)\gamma^\beta,
\end{eqnarray}
\end{subequations}
where $\Lambda(p)=(\,/\!\!\!p+m_c)/(p^2-m_c^2)$.

There are 6 more Feynman diagrams that can be obtained from
the two amplitudes $\mathcal{M}_1$ and $\mathcal{M}_2$
by exchanging two charm quarks and two antiquarks, respectively, as
\begin{equation}
\begin{array}{ll}
 \mathcal{M}_3=-P_{1\leftrightarrow 3}\mathcal{M}_1,
&\mathcal{M}_4=-P_{1\leftrightarrow 3}\mathcal{M}_2,
\\
 \mathcal{M}_5=-P_{2\leftrightarrow 4}\mathcal{M}_1,
&\mathcal{M}_6=-P_{2\leftrightarrow 4}\mathcal{M}_2,
\\
 \mathcal{M}_7=+P_{1\leftrightarrow 3}P_{2\leftrightarrow 4} \mathcal{M}_1,
&\mathcal{M}_8=+P_{1\leftrightarrow 3}P_{2\leftrightarrow 4} \mathcal{M}_2.
\end{array}
\label{eq:amp3-8}
\end{equation}
where $P_{i\leftrightarrow j}$ is the operator exchanging two particles
with momentum indices $p_i$ and $p_j$ shown in Fig.~\ref{fig1}.
The signs of $\mathcal{M}_3$ through $\mathcal{M}_8$ in
Eq.~(\ref{eq:amp3-8}) are determined by the antisymmetricity 
of Fermi statistics in exchanging identical fermions among 
the final-state particles.

The total cross section for the process is expressed as
\begin{equation}
d\sigma=\frac{1}{2s} \overline{\sum}\left|\mathcal{M}\right|^2
\frac{d\Phi_4}{(2!)^2},
\label{eq:sig}
\end{equation}
where $\mathcal{M}=\sum_{i=1}^8 \mathcal{M}_i$, and
the factor $(2!)^2$ in Eq.~(\ref{eq:sig}) is divided in order
to avoid double-counting of identical final-state particles.
The summation notation $\overline{\sum}$ in Eq.~(\ref{eq:sig}) stands for 
averaging over initial spin states and summation over 
final color and spin states.
The four-body phase-space element $d\Phi_4$ in Eq.~(\ref{eq:sig}) 
can be parametrized by
\begin{equation}
d\Phi_4=\frac{dm^2_{13} dm^2_{24}}{(2\pi)^8}
        \cdot
        \frac{|\bm{P}|d\Omega}{4\sqrt{s}}
        \cdot
        \frac{|\bm{p}_1^*|d\Omega^*_{13}}{4m_{13}}
        \cdot
        \frac{|\bm{p}_2^*|d\Omega^*_{24}}{4m_{24}},
\label{eq:ps}
\end{equation}
where $m_{ij}$ is the invariant mass of $p_i+p_j$ and $d\Omega^*_{ij}$ 
is the solid angle element of $p_i+p_j$ in the rest frame of $p_i+p_j$.
Their physical regions are
$2m_c<m_{13}<\sqrt{s}-2m_c$ and $2m_c<m_{24}<\sqrt{s}-m_{13}$.
The three-momenta $\bm{p}_1^*$ of $p_1$
and $\bm{p}_2^*$ of $p_2$ are defined in $p_1+p_3$ and $p_2+p_4$ rest
frames, respectively. The three-momentum $\bm{P}$ and
solid angle element $d\Omega$ are for $p_1+p_2$
in the $e^+e^-$ c.m. frame.
Integrating the differential cross section (\ref{eq:sig})
over the phase space (\ref{eq:ps}),
we get the total cross section for $e^+e^-\to c\bar{c}c\bar{c}$.
We compute the 
$\overline{\sum}|\mathcal{M}|^2$ in Eq.~(\ref{eq:sig})
using REDUCE~\cite{reduce} 
and carry out the phase-space integral in Eq.~(\ref{eq:sig})
making use of the adaptive Monte Carlo routine VEGAS~\cite{vegas}.
As a check, we carry out the same calculation using 
CompHEP~\cite{Boos:2004kh}.
Our analytic result for $\overline{\sum}|\mathcal{M}|^2$ and
numerical values for the total cross section agree with
those obtained by using CompHEP.

\begin{figure}
\includegraphics[width=8cm]{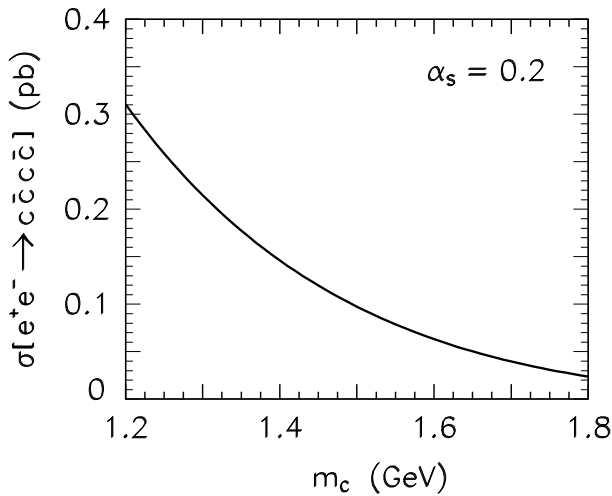}
\caption{\label{fig2}
Total cross section $\sigma(e^+ e^-\to c\bar{c}c\bar{c})$ 
at $\sqrt{s}=10.6$~GeV in pb as a function of $m_c$,
where $\alpha=1/137$ and $\alpha_s=0.2$.
}
\end{figure}
\begin{figure}
\includegraphics[width=8cm]{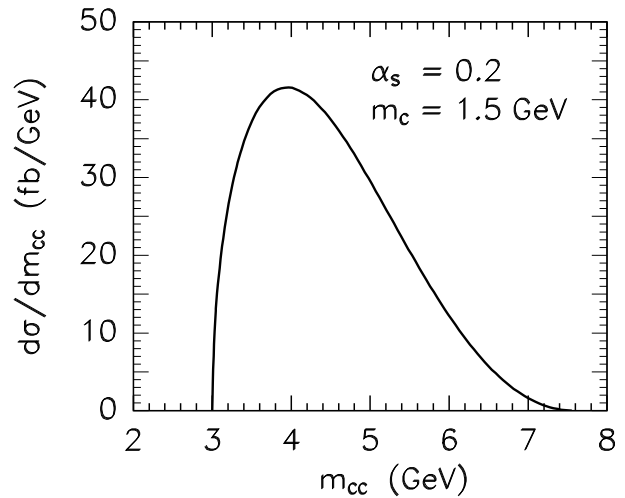}
\caption{\label{fig3}
Differential cross section $d\sigma/dm_{cc}$ in fb/GeV with respect to
the invariant mass $m_{cc}=m_{13}$
of $cc$ for $e^+ e^-$ annihilation into $c\bar{c}c\bar{c}$,
where $m_c=1.5$~GeV,  $\alpha=1/137$, and $\alpha_s=0.2$.
Physical range of the $m_{cc}$ is from
$2m_c$ to $\sqrt{s}-2m_c$.  The area under the curve is 
the integrated cross section 97~fb.
}
\end{figure}

Our predictions for the inclusive four-charm-hadron cross sections 
in $e^+e^-$ annihilation at $\sqrt{s}=$10.6~GeV depending on the
charm-quark mass $m_c$ is  shown in Fig.~\ref{fig2}. 
The cross section for 
$e^+e^-\to c\bar{c}c\bar{c}$ is very sensitive to the value of
$m_c$. For $\alpha=1/137$,  $\alpha_s=0.2$, $m_c=$1.5~GeV
$\sigma(e^+e^-\to c\bar{c}c\bar{c})=$97~fb. 
The cross section varies from 
$0.31$~pb at $m_c=$1.2~GeV to $24$~fb at $m_c=$1.8~GeV.
The cross section decreases as $m_c$ increases mainly because 
available phase space shrinks. If one can increase the c.m. energy 
of the $e^+e^-$, 
the $m_c$ dependence will decrease. 
In previous analyses for exclusive two-charmonium production cross
sections, the next-to-leading order pole mass $m_c=1.4\pm 0.2$~GeV has 
been used for the $m_c$~\cite{Braaten:2002fi,Bodwin:2002fk,Bodwin:2002kk}.
However, the cross section for $e^+e^-\to c\bar{c}$ is
not sensitive to the charm-quark mass $m_c$.
The lowest-order cross section of order $\alpha^2$ is
$\sigma(e^+e^-\to c\bar{c})=1.0$~nb with relative errors of 
$3\times 10^{-3}$ for $m_c=1.5\pm 0.3$~GeV at 
$\sqrt{s}=10.6$~GeV. 
In Ref.~\cite{Berezhnoy:2003hz}
the total cross section for $e^+e^-\to c\bar{c}c\bar{c}$ 
at $\sqrt{s}=10.6$~GeV is predicted.  If we use
the input parameters $\alpha_s=0.24$ and  $m_c=1.4$~GeV
given in Ref.~\cite{Berezhnoy:2003hz}, we get
0.210~pb, which is different from 
the prediction $0.237$~pb given in Ref.~\cite{Berezhnoy:2003hz} by
about 13\%.

In Fig.~\ref{fig3} we show the differential cross section with respect
to the invariant mass of $cc$. This is the prediction for 
$d\sigma(e^+e^-\to cc+X)/dm_{cc}$ in leading order in $\alpha_s$.
Experimentally, this differential cross section can be compared with
the $\sum_{H,H'}d\sigma(e^+e^-\to HH'+X)/dm_{HH'}$, where $H$ and $H'$
are charm hadrons, which do not include anticharm.

Finally, we estimate the number of four-charm-hadron events that could be
detected by the Belle Collaboration. 
Production
rate for baryonic states such as $\Lambda_c$ will be small and we do not 
include the contribution in the following rough estimate. 
Based on heavy-quark spin symmetry, the relative rates for 
a $c$ quark fragmenting into the charm mesons are 
$D^+:D^0:D^{+*}:D^{0*}= 1:1:3:3$. Because the two spin-triplet states decay 
into spin-singlet states by 100\% with branching fractions
Br[$D^{+*}\to D^0 \pi^+$]=70\%,
Br[$D^{+*}\to D^+ \pi^0$]=30\%,
Br[$D^{0*}\to D^0 \pi^0$]=62\%, and
Br[$D^{0*}\to D^0 \gamma$]=38\%,
we may only consider charm meson pairs made of either $D^+$ or $D^0$.
Resulting fragmentation probabilities are approximately 
$P[c\to D^++X]\approx\frac{1}{4}$ and
$P[c\to D^0+X]\approx\frac{3}{4}$, respectively.
Therefore, 
$\sigma[e^+e^-\to D^+D^++X]\approx\frac{1}{16}\sigma[e^+e^-\to cc+X]$,
$\sigma[e^+e^-\to D^+D^0+X]\approx\frac{6}{16}\sigma[e^+e^-\to cc+X]$, and
$\sigma[e^+e^-\to D^0D^0+X]\approx\frac{9}{16}\sigma[e^+e^-\to cc+X]$.
The detection rate will suffer losses from branching fractions 
Br$[D^+ \to K^- \pi^+ \pi^+]=9.2$\% and
Br$[D^0 \to K^- \pi^+      ]=3.8$\%,
and detection acceptance/efficiency $\approx 80$\% 
for each charged particle in the decay products of $D^+$ or $D^0$.
With $\sigma[e^+e^-\to cc+X]\approx 0.1$~pb and 
current integrated luminosity $\mathcal{L}\approx 300$~fb$^{-1}$
we expect roughly 30 events will be detected by the Belle detector.
Even if we consider the uncertainties from 
$\alpha_s$ and $m_c$ in our prediction,
we expect at least about 10 events will be detected by the Belle Collaboration.
If there is a large QCD corrections, the number of events will be increased
into several hundreds.

In summary, we have calculated the cross section for $e^+ e^-$
annihilation into $c\bar{c}c\bar{c}$.  Assuming quark-hadron duality,
the cross section for the inclusive four charm hadrons is predicted
to be about $0.1$~pb. The comparison of this prediction with the
measured cross section for the four charm hadrons
at $B$-factories will provide a strong constraint in determining 
the origin of the large discrepancy between prediction and Belle data
for exclusive $J/\psi+\eta_c$ production in $e^+e^-$ annihilation at
$\sqrt{s}=10.6$~GeV. The measurement will also provide a useful information
in explaining large cross section for $J/\psi+c\bar{c}+X$ 
measured by the Belle Collaboration~\cite{Abe:2002rb}
compared to the NRQCD predictions~\cite{Jpsi-X}.

\begin{acknowledgments}
We would like to thank Eric Braaten and Geoff Bodwin for valuable 
suggestions in improving this manuscript.
We also thank Hongjoo Kim, Shinwoo Nam, and Eunil Won for providing us with
useful experimental information.
PK is supported in part by BK21 Haeksim Program, by
KOSEF through CHEP at Kyungpook National University.
JL is supported by a Korea Research Foundation Grant(KRF-2004-015-C00092).
\end{acknowledgments}



\begin{thebibliography}{}

\bibitem{Abe:2002rb}
K.~Abe {\it et al.}  [BELLE Collaboration],
Phys.\ Rev.\ Lett.\  {\bf 89}, 142001 (2002).

\bibitem{Braaten:2002fi}
E.~Braaten and J.~Lee,
Phys.\ Rev.\ D {\bf 67}, 054007 (2003)
[arXiv:hep-ph/0211085].

\bibitem{Liu:2002wq}
K.~Y.~Liu, Z.~G.~He and K.~T.~Chao,
Phys.\ Lett.\ B {\bf 557}, 45 (2003)
[arXiv:hep-ph/0211181].


\bibitem{BBL}
G.~T.~Bodwin, E.~Braaten, and G.~P.~Lepage,
	Phys.\ Rev.\ D {\bf 51}, 1125 (1995);
{\bf 55}, 5853(E) (1997).

\bibitem{Brodsky:1985cr}
S.~J.~Brodsky and C.-R.~Ji,
Phys.\ Rev.\ Lett.\  {\bf 55}, 2257 (1985).


\bibitem{Brodsky:1981kj}
S.~J.~Brodsky and G.~P.~Lepage,
Phys.\ Rev.\ D {\bf 24}, 2848 (1981).

\bibitem{pQCD}
S.~J.~Brodsky, C.~-R.~Ji, and J.~Lee,
in preparation.

\bibitem{Bodwin:2002fk}
G.~T.~Bodwin, J.~Lee and E.~Braaten,
Phys.\ Rev.\ Lett.\  {\bf 90}, 162001 (2003)
[arXiv:hep-ph/0212181].

\bibitem{Bodwin:2002kk}
G.~T.~Bodwin, J.~Lee, and E.~Braaten,
Phys.\ Rev.\ D {\bf 67}, 054023 (2003)
[arXiv:hep-ph/0212352].

\bibitem{Brodsky:2003hv}
S.~J.~Brodsky, A.~S.~Goldhaber and J.~Lee,
Phys.\ Rev.\ Lett.\  {\bf 91}, 112001 (2003)
[arXiv:hep-ph/0305269].

\bibitem{Abe:2004ww}
K.~Abe {\it et al.}  [Belle Collaboration],
arXiv:hep-ex/0407009.

\bibitem{Hagiwara:2003cw}
K.~Hagiwara, E.~Kou and C.~F.~Qiao,
Phys.\ Lett.\ B {\bf 570}, 39 (2003)
[arXiv:hep-ph/0305102].

\bibitem{Brambilla:2004wf}
N.~Brambilla {\it et al.},
arXiv:hep-ph/0412158.

\bibitem{reduce}
A.~C.~Hearn, \textit{REDUCE User's Manual} v. 3.7 (The RAND Corporation,
Santa Monica, 1999) (Email:reduce@rand.org).
\bibitem{vegas}
G.~P.~Lepage, J. Comput. Phys. \textbf{27}, 192 (1978).
\bibitem{Boos:2004kh}
E.~Boos {\it et al.}  [CompHEP Collaboration],
Nucl.\ Instrum.\ Meth.\ A {\bf 534}, 250 (2004)
[arXiv:hep-ph/0403113].

\bibitem{Berezhnoy:2003hz}
A.~V.~Berezhnoy and A.~K.~Likhoded,
Phys.\ Atom.\ Nucl.\  {\bf 67}, 757 (2004)
[Yad.\ Fiz.\  {\bf 67}, 778 (2004)]
[arXiv:hep-ph/0303145].

\bibitem{Jpsi-X}
P.~Cho and A.~K.~Leibovich,
Phys.\ Rev.\ D {\bf 54}, 6690 (1996);
F.~Yuan, C.-F.~Qiao, and K.-T.~Chao,
Phys.\ Rev.\ D {\bf 56}, 1663 (1997);
Phys.\ Rev.\ D {\bf 56}, 321 (1997);
S.~Baek, P.~Ko, J.~Lee, and H.~S.~Song,
J.\ Korean Phys.\ Soc.\  {\bf 33}, 97 (1998)
[arXiv:hep-ph/9804455];
G.~A.~Schuler,
Eur.\ Phys.\ J.\ C {\bf 8}, 273 (1999).

\end{thebibliography}
\end{document}